\documentclass[12pt,preprint]{revtex4} 

\usepackage{graphics}
\usepackage{epsfig}
\usepackage{amsmath}
\usepackage{amssymb}

\begin{document}

\title{ Modulation control and spectral shaping of optical fiber supercontinuum generation in the picosecond regime}

\author{Go\"{e}ry Genty}
\affiliation{Tampere University of Technology, Institute of Physics,
Optics Laboratory,\\ FIN-33101 Tampere, Finland.}

\author{John M. Dudley}
\affiliation{Institut FEMTO-ST, D\'{e}partement d'Optique P. M. Duffieux,\\ CNRS UMR 6174, Universit\'{e} de Franche-Comt\'{e}, \\
25030 Besan\c{c}on, France.}

\author{Benjamin J. Eggleton}
\affiliation{CUDOS ARC Centre of Excellence, School of Physics, University of Sydney, NSW 2006 Australia.
\vskip 10mm}

\begin{abstract}
\noindent Numerical simulations are used to study how fiber supercontinuum generation seeded by picosecond pulses can be actively controlled through the use of input pulse modulation.  By carrying out multiple simulations in the presence of noise, we show how tailored supercontinuum Spectra with increased bandwidth  and improved stability can be generated using an input envelope modulation of appropriate frequency and depth.  The results are discussed in terms of the non-linear propagation dynamics and pump depletion.

\end{abstract}

\maketitle

\section{Introduction}
\label{intro}

Following its first observation by Ranka \emph{et al.} in 2000 \cite{Ranka-2000a}, supercontinuum (SC) generation in photonic crystal fiber has remained a subject of intense research.  Motivated by important applications in precision frequency metrology, initial effort focussed on developing a clear physical understanding of the underlying mechanisms and noise properties of SC generation seeded by femtosecond pump pulses, a regime which is now very well-understood \cite{Husakou-2001,Gaeta-2002,Corwin-2003a,Dudley-2006}.
Research is now shifting towards detailed studies of the dynamical properties for cases where spectral broadening is initiated in the so-called ``long pulse'' regime, using picosecond to nanosecond pulses, or even a continuous wave pump.  In fact, under these conditions and pumping in the anomalous dispersion regime, the SC spectral broadening has been shown to be associated with a very rich range of dynamical behavior, involving spontaneous pulse break up due to modulation instability (MI) followed by the propagation and interaction between very large numbers of ejected soliton pulses \cite{Dudley-2006,Islam-1989a,Islam-1989b,Vanholsbeeck-2005,Kutz-2005,Frosz-2006,Demircan-2007,Cumberland-2008}.

Recent results have extended this research even further, identifying significant links with
other areas of nonlinear physics.  Particular insight into the mechanism underlying SC fluctuations in the long pulse regime was provided by Solli \emph{et al.} who used a novel real-time detection technique to directly quantify the statistics of picosecond SC shot-to-shot noise \cite{Solli-2007}.  This work was significant in showing that the SC fluctuations led to the generation of ``optical rogue waves,'' statistically rare extreme red-shifted Raman solitons on the long wavelength edge of the SC spectrum.  Further numerical analysis of these fluctuations in Ref. \cite{Dudley-2008} showed explicitly that the rogue soliton statistics exhibit non-gaussian extreme-value characteristics.  In fact, non-gaussian statistics due to soliton collisions in the presence of Raman scattering had been studied earlier in the context of multichannel fibre communications systems \cite{Menyuk-1995,Abdullaev-1996,Georges-1996a,Falkovich-2001a,Derevyanko-2003,Chung-2005a,Peleg-2007a,Peleg-2007b}, but possible links with the soliton dynamics of supercontinuum generation \cite{Islam-1989a,Islam-1989b} were not explored.  Related studies of a fundamental nature in the long pulse regime have applied results from weak turbulence theory to describe the initial spectral broadening as a ``sea of solitons," and the associated semiclassical scattering problem has been shown to analytically model experimentally-measured SC spectra \cite{Korneev-2008}.

From an applications perspective, controlling SC bandwidth and stability is very important and indeed,  guidelines for broadband and low noise SC generation using sub-50~fs  femtosecond pulses are well-known \cite{Dudley-2006}.  Recent studies in the femtosecond regime have also shown how the generated SC spectra can be controlled by varying the input conditions in a more sophisticated manner, for example using dual frequency femtosecond pumping to induce cascaded four wave mixing \cite{Tuerke-2008} or using femtosecond envelope modulation to influence the Raman soliton noise properties \cite{Efimov-2008}.  In the long pulse regime, modified initial conditions can also significantly influence the output SC characteristics and stability.  For example, Solli \emph{et al.} numerically demonstrated a correlation between rogue soliton pulse height and a low amplitude localized noise burst on the leading edge of the picosecond pump pulses \cite{Solli-2007}.  This idea was extended in Ref. \cite{Dudley-2008} where numerical simulations were again used to show that modified rogue soliton statistics could be observed using a $\sim$4$\%$ intensity modulation across the full extent of the pulse envelope.  A recent eprint by Solli \emph{et al.} has reported significant experimental results, showing that seed modulation of SC generation at the -30~dB level can also introduce an effective phase transition in the SC stability \cite{Solli-2008}.

The dramatic effect of these modified input conditions illustrate the sensitivity of the initial MI propagation phase of long pulse SC generation to coherent input modulation.
However, the studies described above have considered only the effects of weak perturbations over a limited parameter range, whereas it might be expected that the induced SC dynamics will depend very sensitively on both the frequency and amplitude of any applied modulation.  Our objective in this paper is therefore to examine the potential parameter space of input modulation more extensively, focussing on how both the  SC spectral intensity and stability are modified as a function of the modulation parameters.  A major result is the discovery that certain modulation parameters yield stabilised spectra with Raman soliton peak power statistics transformed from an ``L-shaped'' extreme-value distribution to a near-gaussian distribution with significantly reduced peak power fluctuations below $5\%$.  We also extend previous physical discussions of the SC generation mechanism in the long pulse regime to explicitly include the role of pump depletion.  These results show that considerations of pump depletion dynamics can provide useful insight into the sensitivity of the SC broadening to input pulse modulation.

\section{Numerical Model}
\label{numerics}

Our simulations use a generalized form of the well-known nonlinear Schr\"{o}dinger equation suitable for modeling propagation of broadband unidirectional fields \cite{Dudley-2006,Blow-1989}.  With explicit inclusion of higher-order linear and nonlinear terms, it can be written in the following way:
\begin{multline}
  \frac{\partial{A}}{\partial{z}}+\frac{\alpha}{2}A
    -\sum_{k\geq2}{\frac{\mathrm{i}^{k+1}}{k!}\beta_k \frac{\partial^k A}{\partial T^k} }
    = i\gamma\left(1+i \tau_{\mathrm{shock}}\frac{\partial}{\partial T}\right)\times \\
 \left( A(z,t)\left[ \int_{-\infty}^{+\infty}
   \!\!\! R(T')|A(z,T-T')|^{2}dT'\! + \! i\Gamma_{\mathrm{R}}(z,T)\,\right]\right).
 \label{GNLSE1}
\end{multline}


Here $A(z,t)$ is the field envelope and the $\beta_\mathrm{k}$'s and $\gamma$  are the usual dispersion and nonlinear coefficients.  In our simulations we consider propagation in 25~m of fused silica based optical fibre with zero dispersion around 1055~nm. The dispersion coefficients at a pump wavelength of 1060~nm are:
$\beta_2 =  -0.820 \, \mathrm{ps}^2 \, \mathrm{km}^{-1}$, $\beta_3 =  6.87 \times 10^{-2} \, \mathrm{ps}^3 \,\mathrm{km}^{-1}$, $\beta_4 =  -9.29 \times 10^{-5} \, \mathrm{ps}^4 \,\mathrm{km}^{-1}$, $\beta_5 =  2.45 \times 10^{-7} \, \mathrm{ps}^5 \,\mathrm{km}^{-1}$, $\beta_6 =  -9.79 \times 10^{-10} \, \mathrm{ps}^6 \,\mathrm{km}^{-1}$, $\beta_7 =  3.95 \times 10^{-12} \, \mathrm{ps}^7 \,\mathrm{km}^{-1}$, $\beta_8 =  -1.12 \times 10^{-14} \, \mathrm{ps}^8 \,\mathrm{km}^{-1}$, $\beta_9 =  1.90 \times 10^{-17} \, \mathrm{ps}^9 \,\mathrm{km}^{-1}$, $\beta_{10} = -1.51\times 10^{-20} \, \mathrm{ps}^{10} \,\mathrm{km}^{-1}$ and $\gamma = 0.015 \mathrm{W}^{-1}\mathrm{m}^{-1}$.  The nonlinear response $R(t) = (1-f_\mathrm{R}) (t) + f_\mathrm{R} h_\mathrm{R}(t)$ with $f_\mathrm{R} = 0.18$ includes instantaneous and Raman contributions.  The response $h_\mathrm{R}$ is determined from the experimental fused silica Raman cross-section \cite{Stolen-1989}, but similar results can be obtained using analytic approximations for the Raman term \cite{Agrawal-2006}.  The envelope self-steepening timescale $\tau_\mathrm{shock} = 0.658 $~fs.  Noise is included in the frequency domain through a one photon per mode background, and via the term $\Gamma_\mathrm{R}$ which describes thermally-driven spontaneous Raman scattering \cite{Dudley-2006,Drummond-2001}.  To examine the SC stability characteristics, multiple simulations were carried out in the presence of different random noise seeds, allowing the spectral and temporal structure of the SC to be examined from shot-to-shot.  In this regard, we have found that the spectral shape and coherence properties can be well determined using an ensemble of typically 100 realizations, although for cases when histogram data is calculated,  obtaining sufficient data in the distribution tails requires larger ensembles with $> 1000$ realizations.

We consider an input field of the form:
$A(0,T) = \sqrt{P_0}\, \exp(-t/2\,T_0^2)\,[1 + a_0 \exp(-\mathrm{i}\Omega t)]$, which corresponds to a modulated gaussian pulse envelope with intensity modulation contrast $2a_0/(1+a_0^2)$.  A modulated envelope of this form can be generated through the beating of a pulse with a (Stokes) frequency-shifted replica, and the use of such dual frequency fields has been previously used to generate high-repetition rate pulse trains through induced MI/four wave mixing \cite{Greer-1989,Mamyshev-1991,Dudley-2001}.  We carry out simulations over a range of values of $a_0$ and $\Omega$, adjusting the pump and Stokes sideband amplitude to ensure that the energy remains constant.  With this approach, the individual peak powers of the pump and Stokes pulses are $P_\mathrm{p} = P_0/(1+a_0^2)$ and $P_\mathrm{s} = P_0\, a_0^2/(1+a_0^2)$ respectively.  We choose an input energy of 0.4~nJ which corresponding to
$P_\mathrm{p}= P_0 = 75$~W at zero modulation.

\section{Numerical Simulations}
\label{Results}

\subsection{Illustrative Results}

\begin{figure}
\resizebox{1\textwidth}{!}{
\includegraphics{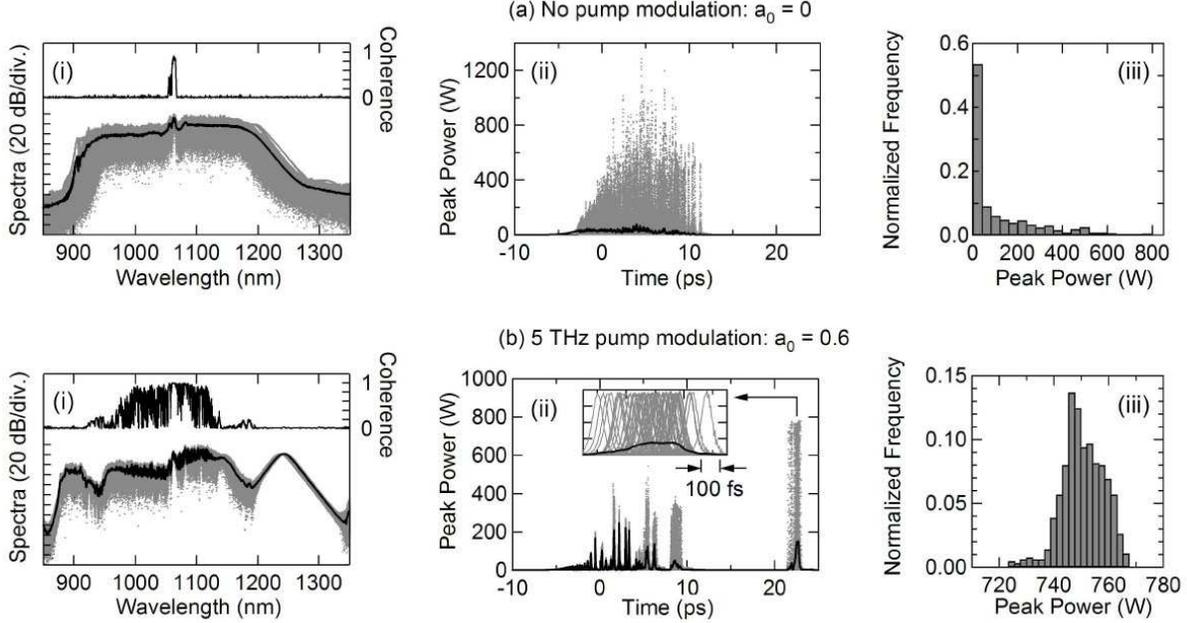}}
\caption{Simulation results for SC induced by pulses with (a) no input modulation and (b) modulation at 5 THz and $a_0=0.6$.  For each case, (i) and (ii) show output spectral and temporal characteristics respectively, superposing individual simulation results (gray traces) with the calculated mean from the ensemble(black line).  (i) also plots calculated coherence on the right axis.   (iii) shows the peak power histogram after spectral filtering at 1180~nm, plotting normalized event frequency such that bar height represents the proportion of data in each bin.   Note that coherence and statistical data is calculated from the full ensemble of 1000 realisations, but for clarity, the gray traces show only 75 realisations.}
\label{Fig1}
\end{figure}

We first consider SC generation in the absence of any envelope modulation ($a_0 = 0$) seeded by pulses with $T_0 = 3$~ps and $P_0 = 75$~W.  At this (anomalous dispersion) pump wavelength, the soliton order $N=(\gamma\,P_0\,T_0^2/|\beta_2|)^{0.5} \approx 111 $. Such a large value is typical of picosecond SC generation, and is expected to yield significant shot-to-shot variation in both the spectral and temporal characteristics \cite{Dudley-2006}. This is illustrated explicitly in Fig.~\ref{Fig1}(a) which shows results from an ensemble of 1000 simulations, plotting the output SC characteristics after 25~m propagation.  Subfigures (i) and (ii) show the output spectral and temporal characteristics respectively, superposing a subset of 75 realisations from individual simulations (gray traces) and also showing the calculated mean from the full ensemble (black line).

Fig.~1 clearly shows significant shot-to-shot fluctuations in both the temporal and spectral intensities.  Fluctuations in the spectral phase across the SC can be seen by plotting the spectral coherence calculated from the ensemble, and this is plotted on the right axis of subfigure (i).  This coherence function corresponds to the fringe visibility at zero path difference in a Young's two source experiment performed between independent SC spectra, and is a widely-used measure of SC stability \cite{Dudley-2002a}.  Of significance here is that the shot-to-shot fluctuations essentially lead to zero spectral coherence (spectral phase stability) across the full bandwidth of the SC, except for a narrow wavelength range in the vicinity of the pump.   Such instability in both the temporal and spectral properties is a well-known characteristic of SC generation using picosecond pulses, arising because spectral broadening in this regime is seeded from an ultrafast modulation that develops on the pulse envelope from noise-induced MI.  As the envelope subsequently breaks up into individual soliton pulses with further propagation, the random nature of the initial modulation introduces significant shot-to-shot differences in the subsequent Raman soliton dynamics and frequency shifts.

In this regard, the spectral plot in subfigure(i) of Fig. 1(a) illustrates the small number of events that are associated with large wavelength shifts and the generation of ``rogue'' soliton pulses \cite{Solli-2007,Dudley-2008}. These particular events can be isolated applying the technique developed in Ref. \cite{Solli-2007} using a spectral filter to select components above a particular wavelength on the long-wavelength edge.  Fourier transformation yields a series of ultrashort pulses of varying power depending on the position of the filter relative to the SC spectral structure.  The statistical frequency distribution of the pulse peak power then readily reveals the presence of rare high peak-power events that have been fully captured because of their extreme long wavelength shifts.  Subfigure (iii) shows the corresponding histogram obtained from the data using a filter at 1180~nm illustrating the heavy-tailed ``L-shaped'' probability distribution which is typical of extreme value processes.


As discussed in the introduction above, the results in Ref.~\cite{Dudley-2008} showed how a weak envelope modulation could significantly modify the rogue soliton statistics.  In this study, we have examined the effect of such input modulation for the case of higher modulation depths and have found that modulation depths above $50\%$ in fact yield a very significant degree of control into the output SC  properties compared to the case where the SC develops from noise.  This is conveniently illustrated by considering a specific case, and Fig.~1(b) shows the output SC characteristics for an input field consisting of a pump at 1060~nm and a Stokes pulse of the same duration, but frequency-shifted by 5~THz (i.e: at a wavelength of 1079~nm), and with $a_0=0.6$ such that $P_\mathrm{p} = 55$~W and $P_\mathrm{s} = 20$~W.

Significantly, although the peak power at the pump wavelength is reduced compared to the unmodulated case, Fig.~1(b) shows clearly that the effect of this envelope modulation leads to significantly different output characteristics for the spectral, temporal and statistical properties.  For example, subfigure (i) shows an increased overall bandwidth with extended long and short wavelength edges, improved spectral coherence and significantly less variation in the shot-to-shot spectral structure compared to Fig.~1(a).  This leads to a well-isolated Raman soliton peak on the long wavelength edge.  The time domain characteristics in subfigure (ii) also show reduced shot-to-shot variation when compared to Fig.~1(a) and clearly illustrate the localized temporal structure of the Raman soliton around a time coordinate of 24~ps.  However, the inset shows that although each Raman soliton has a temporal duration of $\sim80$~fs, the residual wavelength fluctuations result in significant temporal jitter in the soliton position of $\sim$1~ps.  It is this temporal jitter that leads to the near zero spectral coherence in subfigure (i) in the vicinity of the Raman soliton wavelength.  Nonetheless, when compared to the spontaneously generated SC in Fig.~1(a) the input modulation yields a remarkable improvement in the SC stability properties.  This is also seen in subfigure (iii) which shows the peak power histogram after long wavelength filtering at 1180~nm.  Here we see that instead of an L-shaped distribution indicative of a small number of extreme red-shifted rogue events, we obtain  a significantly more localized gaussian-like distribution.  Indeed, the fluctuation in the filtered soliton pulse peak power about the mean for these parameters is only $\approx5\%$.

\subsection{Interpretation and Discussion}

To interpret these results physically, it is useful to consider how the SC characteristics vary a wider range of  modulation frequencies.  To this end, Fig.~2 shows results for $a_0=0.6$ but over a 0--20 THz span of modulation frequencies.   The results show (a) the mean output spectra and (b) the associated wavelength-dependent degree of coherence calculated from the ensemble.  Here we plot the results using a false color representation with wavelength as the horizontal axis and modulation frequency on the vertical axis.  Note that these spectral characteristics calculated over an ensemble generalise  similar single-shot noise-free results used in previous work \cite{Dudley-2008}, and provide more realistic predictions of the spectral structure and stability properties that could be expected in experiments.

From these results, we see that the output spectral structure clearly exhibits significant dependence on modulation frequency.  Indeed, it can be seen that  the particular frequency of the input modulation introduces a remarkable degree of control into the SC generation process, allowing both extension and reduction of the output bandwidth when compared to the unmodulated case.  The dependence of the output spectral structure on modulation frequency is complex, but useful insight can be obtained by comparing the results seen in Figs. 2(a) and (b) with the frequency-dependent gain curve describing the growth of a Stokes wave component due to MI and Raman processes.  To this end, we plot in Fig.~2(c) the gain curve describing the coupling between MI and Raman gain \cite{Vanholsbeeck-2003},calculated for the undepleted pump wave of power $P_\mathrm{p}$ at the fiber input.  Note that the curve is more complex than a mere superposition of the separate MI and Raman processes, but nonetheless shows a clear MI gain peak around 8~THz and a distinct shoulder around the 13.2~THz peak of Raman gain in fused silica.

\begin{figure}[h!]
\resizebox{1\textwidth}{!}{%
\includegraphics{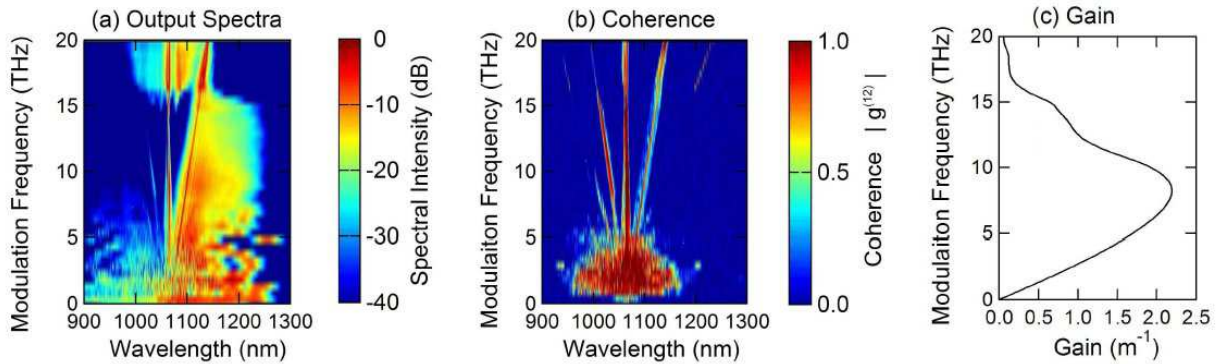}
}
\caption{(a) Density plot of output spectral intensity after 25~m propagation for $a_0=0.6$ as a function of applied modulation frequency (vertical axis).  (b) Corresponding degree of spectral coherence.  (c) Calculated mixed MI-Raman gain curve.}
\label{Fig2}
\end{figure}

From  inspection of Fig. 2(a), we can identify three broad ranges for the input modulation frequency that present qualitatively different output characteristics.  Firstly, for modulation frequencies less than $\sim 8$~THz, the output spectra extend on both the long and short wavelength side of the pump and exhibit soliton and dispersive wave structure.  These characteristics illustrate the importance of soliton dynamics in this regime, arising because of the reshaping of the input pulse envelope into a train of high contrast soliton train during the initial MI phase of propagation.  Significantly, we note that the values of modulation frequency detunings where we observe these dynamics are less than the frequency of maximum MI gain.   Although perhaps surprising, this can be understood from the fact that we are in a strong conversion regime, and thus frequency of maximum MI gain dynamically shifts towards lower values with propagation as pump depletion becomes significant.

In fact, previous analytical studies of this process under CW excitation using a truncated sideband model to describe coherent pump-energy exchange have predicted maximum integrated gain with an initial modulation frequency of around half the frequency of peak MI gain calculated using the initial value of undepleted pump power \cite{Cappellini-1991b}.  Although the dynamics of the regime considered here is more complex (involving multiple sidebands and Raman scattering), such an analysis is useful for the physical interpretation of these results.  In particular, we can interpret the improved coherence properties of the generated SC seen in Fig.~2(b) [see also Fig. 1{b)] as arising from the reduced influence of spontaneous MI relative to coherent exchange between the generated sidebands.

Secondly, as the modulation frequency increases into the range 8--15~THz, Fig.~ 2(a) now shows a predominantly red-shifted output SC spectrum where long wavelength soliton structure is no longer clearly apparent.  This behaviour can be readily understood from the fact that the dynamically decreasing frequency shift of the MI gain curve due to pump depletion means that increased initial modulation frequency lies outside the region of significant MI gain after a very short propagation distance and thus the dynamics exhibit reduced coherence as reflected in Fig.~2(b). In this regime, Raman amplification plays a more significant role and there is a predominantly red-shifted output spectrum.  A more detailed analysis on a shot-to-shot basis still shows the development of red-shifted solitons from the Raman-amplified Stokes side-band, but these are not distinctly resolved in the mean spectrum.

Finally, for frequencies exceeding $\sim 15$~THz, the input modulation lies outside the regime of significant MI and Raman gain.  In this case, we see the growth of low amplitude spontaneous MI sidebands around 7~THz from the pump so that this regime is essentially similar to the incoherent and unstable case of an unmodulated pump in Fig. 1(a).   A detailed consideration of this regime would, however, need to take into account energy exchange between the pump and both initial and spontaneously generated sidebands with propagation.

\begin{figure}[ht]
\resizebox{1\textwidth}{!}{%
\includegraphics{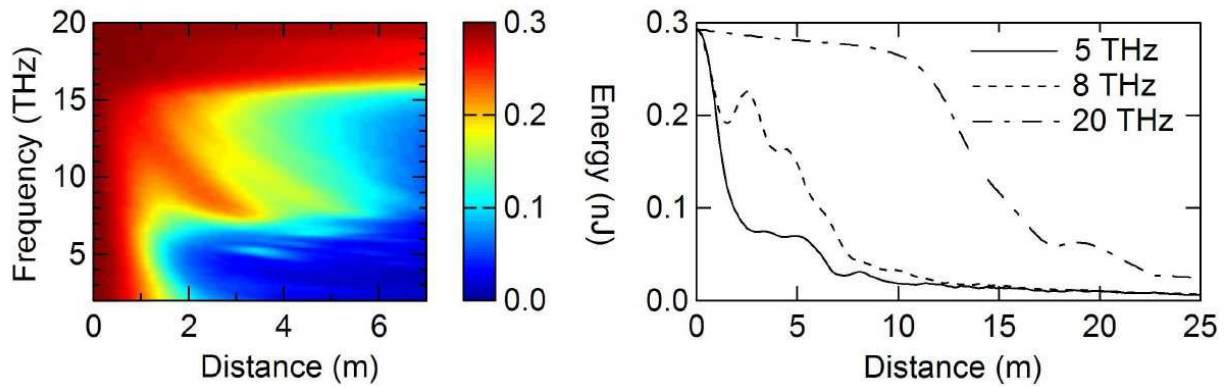}
}
\caption{Energy over 1~THz bandwidth at the pump wavelength as a function of propagation distance.  Left: false color representation for modulation frequencies over the range 1--20~THz up to a propagation distance of 7~m.  Right: pump energy over 25~m propagation for selected modulation frequencies as shown.}
\label{Fig3}
\end{figure}

Fig. 3 and 4 illustrate additional aspects of the dynamics for this case of a 5~THz modulation and with $a_0=0.6$.  Specifically, Fig.~3 highlights the importance of pump depletion in the initial phase of propagation, plotting the energy at the pump wavelength as a function of propagation distance, and for modulation frequencies over the range 1--20 THz.  Note in calculating the energy at the pump wavelength we integrate over a 1 THz bandwidth; this is more than 10 times the initial bandwidth, and allows the depletion of pump energy through gain dynamics to be fully captured while covering a bandwidth sufficient to include the broadening of the pump itself due to self-phase modulation.

The left subfigure shows results using a false color representation  up to a propagation distance of 7~m, whilst the right subfigure plots the pump energy over 25~m propagation for selected modulation frequencies as shown.  Although the detailed frequency and distance dependence of the pump depletion is complex, it is nonetheless clear that maximum pump depletion occurs at an earlier stage of the propagation for modulation frequencies less than 8 THz, correlating with coherent MI gain dynamics and the more distinct observed soliton features in the spectra shown in Fig.~2.

\begin{figure}[ht]
\resizebox{1\textwidth}{!}{%
\includegraphics{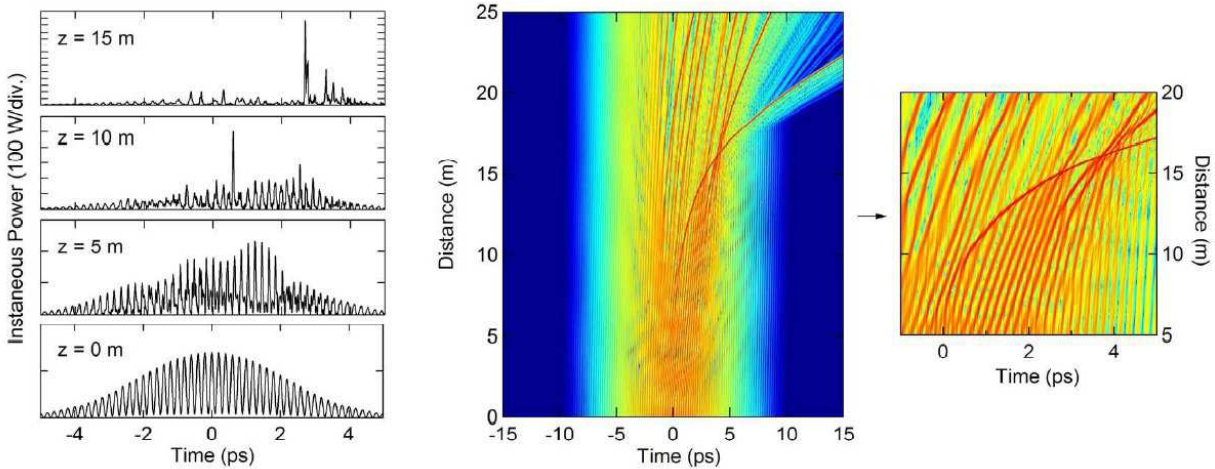}
}
\caption{Dynamics for the spectral broadening in Fig. 1(b)  with modulation frequency of 5~THz and amplitude $a_0 = 0.6$.  Left: line plots of temporal intensity profile at selected distances.  Right: false colour plot of the intensity profile evolution, with a detailed view showing how the largest soliton component collides with and extracts energy from other soliton pulses on the pulse envelope.}
\label{Fig4}
\end{figure}

The particular modulation frequency of 5~THz considered in Fig.~1 leads to a very clear and distinct soliton component on the long wavelength edge of the output spectra, and Fig. 4 considers the corresponding time-domain dynamics in more detail.  This figure shows how the evolution of the initially modulated envelope leads to the ejection of one particular high-power soliton pulse (after a distance of around 7 m) that extracts energy from the other soliton like pulses on the envelope, before clearly separating from the residual envelope in the time domain due to its significantly different group velocity.  As mentioned in the introduction, the exchange of energy between colliding solitons in the presence of the Raman effect has been the subject of much previous research \cite{Frosz-2006,Peleg-2007a} and it is likely that this process plays a key role in the statistical excitation of rare rogue wave events in the case where supercontinuum generation arises from noise \cite{Solli-2007,Dudley-2008}.  The significance of the results shown in Fig.~4 is that they suggest that the study of the propagation dynamics of an initially modulated input may allow this energy exchange process to be studied under controlled (rather than statistical) conditions.

As a final point of discussion, we note that similar qualitative features are observed over a wider range of modulation depths, and Fig.~5 shows false colour density plots showing the modulation frequency dependence of the mean output spectra and associated coherence for different values of the modulation depth.   Although the detailed interpretation of the results must be carried out on a case-by-case basis, we can clearly see that increased modulation depths are associated with improved coherence in the 0--8~THz range where coherent MI gain processes are significant.  This can be readily understood from the fact that the strong conversion regime where coherently seeded MI initiates the SC process is only reached for larger values of the modulation depths, in which case the influence of spontaneous MI is strongly reduced and overall coherence is enhanced.

\begin{figure}
\resizebox{1\textwidth}{!}{%
\includegraphics{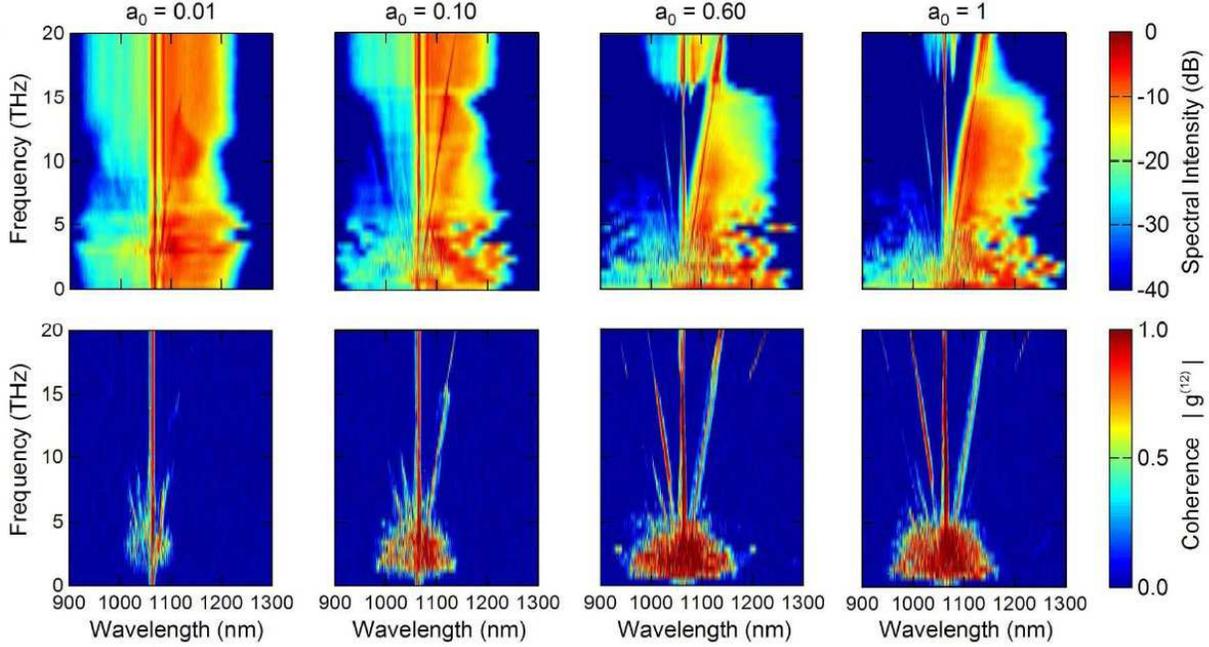}
}
\caption{Top: false color density plots of output spectra as a function of modulation frequency for four values of modulation index as indicated.  Bottom: corresponding plots of the mutual coherence function.}
\label{Fig5}
\end{figure}

\section{Conclusions}
\label{Conclusions}

In this work we have investigated the possibility of controlling the generation of supercontinuum in the long pulse regime through modulation of the input pulse. Our results indicate that depending on the value of the modulation frequency and depth the mean supercontinuum spectral shape can be significantly  modified. In addition, we have shown that, for large values of the modulation depth and modulation frequencies in a range determined by the initial MI-Raman gain, SC spectra with extended bandwidth and improved stability can be generated.  These results may find application in the tailoring of broadband SC spectra for specific applications.

\section{Acknowledgements}

We thank the \emph{Institut Universitaire de France}, the \emph{Agence Nationale de Recherche} (ANR, Projet SOFICARS) and the Academy of Finland for support.  We also acknowledge valuable discussions with members of the French national GDR research network PhoNoMi2 and the European COST299 Action.


\providecommand{\micron}{\ifmmode\mu{\mathrm m}\else$\mu{\mathrm m}$\fi}
  \providecommand{\singleletter}[1]{#1} \providecommand{\hideforsort}[1]{}

\end{document}